\documentclass[]{interact}

\usepackage{epstopdf}
\usepackage{subfigure}

\usepackage[numbers,sort&compress,merge]{natbib}
\bibpunct[, ]{(}{)}{,}{n}{,}{,}

\theoremstyle{plain}

\theoremstyle{definition}

\theoremstyle{remark}

\begin{document}


\title{Motional studies of one and two laser-cooled trapped ions for electric-field sensing applications}

\author{
\name{F.~Dom\'inguez\textsuperscript{a}, M.J.~Guti\'errez\textsuperscript{a}, I.~Arrazola\textsuperscript{b}, J.~Berrocal\textsuperscript{a}, 
J.M.~Cornejo\textsuperscript{a,*}\thanks{$^*$Present address: Institut f\"ur Quantenoptik, Leibniz Universit\"at Hannover, Welfengarten 1, 30167 Hannover, Germany}, J.J.~Del~Pozo\textsuperscript{a}, R.A.~Rica\textsuperscript{a}, S.~Schmidt\textsuperscript{a}, E.~Solano\textsuperscript{b,c}, and D.~Rodr\'iguez\textsuperscript{a,d,**}\thanks{$^{**}$Corresponding author: D.~Rodr\'iguez. Email: danielrodriguez@ugr.es}}
\affil{\textsuperscript{a}Departamento de F\'isica At\'omica, Molecular y Nuclear, Universidad de Granada, 18071, Granada, Spain;
\textsuperscript{b}Department of Physical Chemistry, University of the Basque Country UPV/EHU, Apartado  644, 48080 Bilbao, Spain;
\textsuperscript{c}IKERBASQUE, Basque Foundation for Science, Maria Diaz de Haro 3, 48013 Bilbao, Spain;
\textsuperscript{d}Centro de Investigaci\'on en Tecnolog\'ias de la Informaci\'on y las Comunicaciones,  Universidad de Granada, 18071, Granada, Spain}
}

\maketitle

\begin{abstract}
We have studied the dynamics of one and two laser-cooled trapped $^{40}$Ca$^+$ ions by applying electric fields of different nature along the axial direction of the trap, namely, driving the motion with a harmonic dipolar field, or with white noise. These two types of driving induce distinct motional states of the axial modes; a coherent oscillation with the dipolar field, or an enhanced Brownian motion due to an additional contribution to the heating rate from the electric noise. In both scenarios, the sensitivity of an isolated ion and a laser-cooled two-ion crystal has been evaluated and compared. The analysis and understanding of this dynamics is important towards the implementation of a novel Penning-trap mass-spectroscopy technique based on optical detection, aiming at improving precision and sensitivity.
\end{abstract}


\section{Introduction}
Electronic-based precision sensing in ion traps allows for the accurate determination of the eigenfrequencies of a stored charged-particle from the current it induces in the trap electrodes. This minute current is amplified by means of a tuned circuit operated at liquid-helium temperature, which is attached to several amplification stages or inductively coupled to a DC SQUID \cite{Wine1975,Corn1989,Vand1995}. This method has been implemented at several university laboratories using a single ion, or even two moving in the same magnetron orbit, to perform measurements of the mass-to-charge ratio of the stored particle with relative uncertainties in the order of $10^{-11}$. Applications range from e.g., the determination of the fine structure constant $\alpha ^{-1} = 137.0359922(40)$ using the measurements on the mass doublets $^{133}$Cs$^{3+}$-CO$_2^+$ and $^{133}$Cs$^{3+}$-C$_5$H$_6^+$ \cite{Brad1999}, test of the Einstein's relationship $E = mc^2$ from measurements on $^{29}$Si$^+$-$^{28}$SiH$^+$ and $^{33}$S$^+$-$^{32}$SH$^+$ \cite{Rain2005}, or delivering important contributions in metrology by determining the mass of $^{28}$Si$^+$ for the proposed silicon-ball based S.I.  kilogram \cite{Reds2008}. The technique itself has allowed carrying out the most accurate measurement of the mass of the electron $m_e=0.000 548 579 909 067(14)(9)(2)$\,u using $^{12}$C$^{5+}$ \cite{Sturn2014}, and that of the proton $m_p = 1.007 276 466 583(15)(29)$\,u \cite{Heis2017}. This detection scheme has also been implemented in Penning traps operating at large scale facilities like CERN, where already  in 1999, a comparison between the mass of the proton and the mass of the antiproton was quoted with an uncertainty of $9\times 10^{-11}$ \cite{Gabr1999}. This ratio has been recently improved to $1(69)\times 10^{-12}$, representing the most precise CPT invariance test within the Standard Model \cite{Ulme2015}. \\

\noindent Despite the success of the Induced Image Current (IIC) detection technique, it has not been demonstrated yet with singly-ionized heavy or superheavy elements. Furthermore, short-lived atomic nuclei produced at accelerators might not be practicable with this technique due to the time needed to perform the measurement, which might be considerably larger than the nuclear-decay half life. Besides the developments being carried out in IIC detection at several Penning traps coupled to Radioactive Ion Beam facilities, a different approach, based on using a laser-cooled $^{40}$Ca$^+$ ion{\footnote {For laser-cooling in an ion trap see e.g. \cite{Blat2003}.}} as sensor, to optically detect motional frequencies of the ion of interest, is under development at the University of Granada \cite{Rodr2012}. The final goal will be accomplished in two steps: first by monitoring two ion species in the same Penning trap, in similar way as it was done in a Paul trap for two $^{40}$Ca$^+$ ions \cite{Drew2004}, and second by storing the laser-cooled ion in a different and adjacent trap, following a previous idea described elsewhere \cite{Wine1990}. Monitoring two-ion species in the same trap might allow for fast and sufficiently precise mass measurements on Superheavy Elements produced at GSI and so far not available at the SHIPTRAP facility \cite{Bloc2010,Mina2012}. The second approach of storing the ions in different traps \cite{Rodr2012,Corn2016} will allow probing eigenfrequencies of the ion of interest in the absence of Coulomb interaction, while still continuously monitoring the magnetic field strength through the sensor ion. Since very small oscillation amplitudes can be detected optically \cite{Domi2017}, this might allow reducing the relative mass uncertainty to the level where the results can be utilized for neutrino mass spectrometry (see e.g. \cite{Elis2015}).\\

\noindent In this contribution we characterize analytically the motion of two Doppler-cooled trapped $^{40}$Ca$^+$ ions exposed to external fields of different nature, detailing the description of the dynamics of the two-ion crystal in the presence of white noise and the optical response of the imaging system. The observation of eigenfrequencies in a two-ion crystal through the fluorescence signal from a laser-cooled $^{40}$Ca$^+$ ion is the first approach in our on-going Penning-trap experiment.

\section{The Paul-trap setup}

\begin{figure}[t]
\centering
\includegraphics[width=1.2\linewidth]{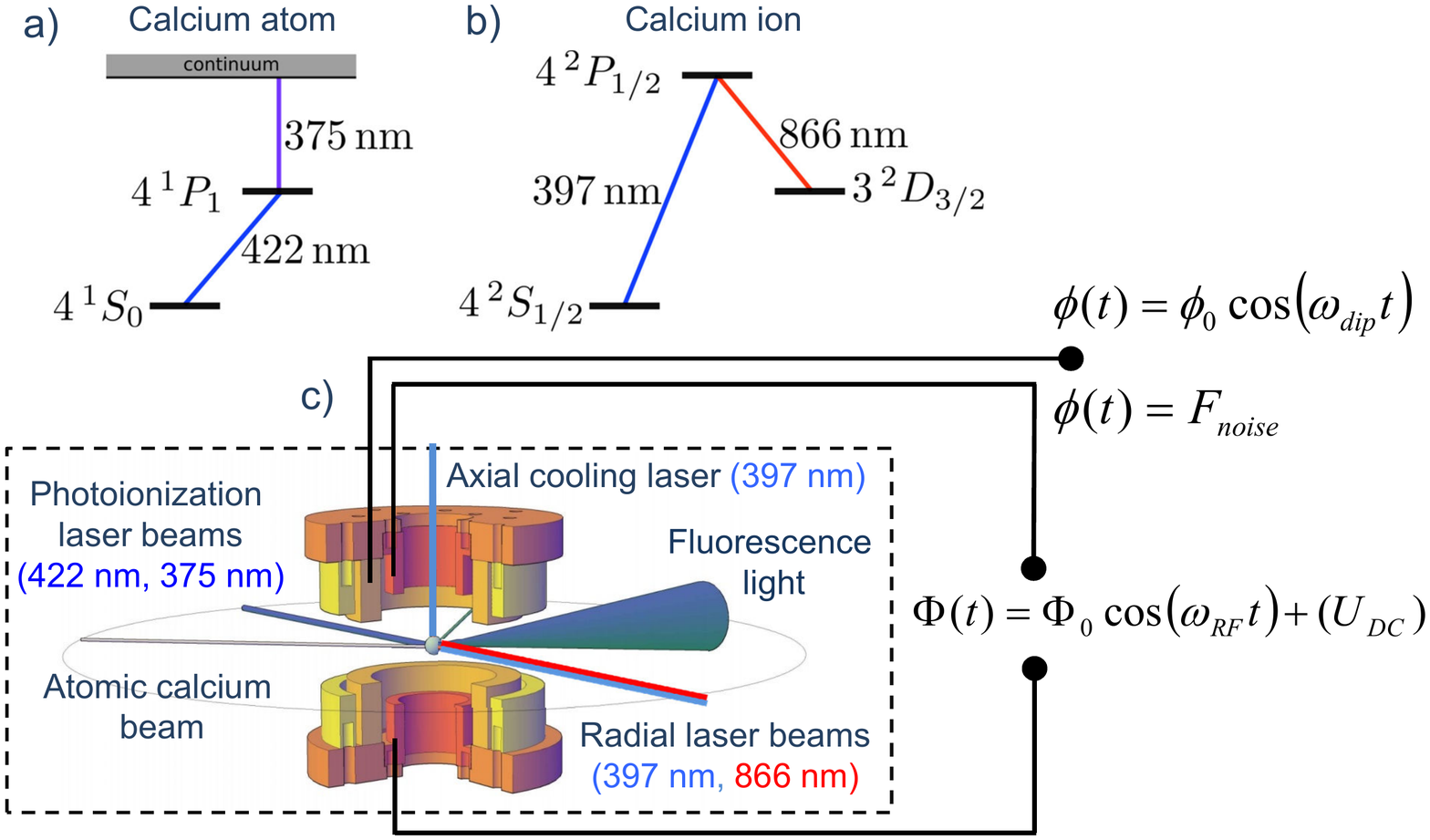}
\vspace{-3.5cm}
\caption{a) Level scheme for the photoionization of Ca atoms, b) relevant level structure for the laser cooling of $^{40}$Ca$^+$, c) three dimensional technical drawing of the open-ring Paul trap used for the experiments reported in this publication. Three laser beams are used for Doppler cooling in the axial (1x 397~nm) and radial (1x 397~nm, 1x 866~nm) directions. Another two lasers are utilized to photoionize the calcium atoms.  The cone in the figure depicts the photons from the fluorescence distribution collected with the optical system. The scheme to apply the different voltages to perform the experiments reported here is also shown.\label{fig_1}}
\end{figure}

The experiments reported in this publication have been carried out using $^{40}$Ca$^+$ ions confined and laser-cooled in an open-ring Paul trap. The experimental set-up was described in detail in Ref.~\cite{Corn2015} and here only the main components are outlined. It is made of two sets of three concentric rings centred on the $z$-axis, built inside a 6-way CF100 vacuum cross operated at a background pressure of $\approx 1.0\times 10^{-10}$~mbar. The $^{40}$Ca$^+$ ions are created inside the trap by two-step photoionization using near-UV diode lasers, one to drive the transition 4$^1$S$_{0}\rightarrow \,$4$^1$P$_{1}$ in the $^{40}$Ca atom ($\lambda = 422$~nm), and another one to bring the electron to the continuum ($\lambda =375$~nm). Figure~\ref{fig_1} shows a cut from a 3D CAD drawing of the Paul trap indicating the directions of the atomic calcium beam together with the laser beams for photoionization, and those utilized to perform Doppler cooling, i.e., to drive the cooling transition 4$^2$S$_{1/2}\rightarrow$4$^2$P$_{1/2}$ with $\lambda =397$~nm, and to pump the metastable 3$^2$D$_{3/2}$ state to 4$^2$P$_{1/2}$ with $\lambda = 866$~nm. The fluorescence light from the cooling transition (397~nm) is collimated using a commercial system, which provides a magnification of $\simeq 6.75(5)$ in the focal plane of an EMCCD (Electron Multiplier Charged Couple Device). An effective pixel size of $\approx 2.4$~$\mu $m has been obtained from the ratio of the minimum distance between two trapped ions due to Coulomb interaction, to the distance in pixels separating the centroids of the fluorescence distributions projected on the axial direction.\\

\noindent The trapping field is generated by a radiofrequency RF signal with frequency $\omega _{\scriptsize{\hbox{RF}}}=2\pi \times 1.47$~MHz, and voltage $V_{\scriptsize{\hbox{RF}}}=730$~V$_{\scriptsize{\hbox{pp}}}$ ($2 \times \Phi _0$), and an electrostatic potential $U_{\scriptsize{\hbox{DC}}}=-11.5$~V to the same electrodes, as shown in Fig.~\ref{fig_1}. The secular frequency is given by 
\begin{equation}
\omega _{u}=\frac{\omega _{\scriptsize{\hbox{RF}}}}{2}\sqrt{a _{u}+\frac{q  _{u}^{\scriptsize{\hbox{2}}}}{2}},   \qquad u=r,z
\end{equation}
where $q  _{u}$ and $a_{\scriptsize{\hbox{u}}}$ are the so-called Mathieu parameters \cite{Ghos1995}. The trap (secular) frequency in the axial direction $\omega _{\scriptsize{\hbox{z}}}$ is about $2\pi \times 80$~kHz, and for these values of $V_{\scriptsize{\hbox{RF}}}$ and $U_{\scriptsize{\hbox{DC}}}$, $\omega _{\scriptsize{\hbox{z}}}<\omega _{\scriptsize{\hbox{r}}}$. The Mathieu parameter $q _{\scriptsize{\hbox{z}}} \simeq 0.25$,  is within the range where the adiabatic approximation is valid~\cite{Ghos1995}.

\section{Experimental results}

\begin{figure}[t]
\centering
\includegraphics[width=1.1\linewidth]{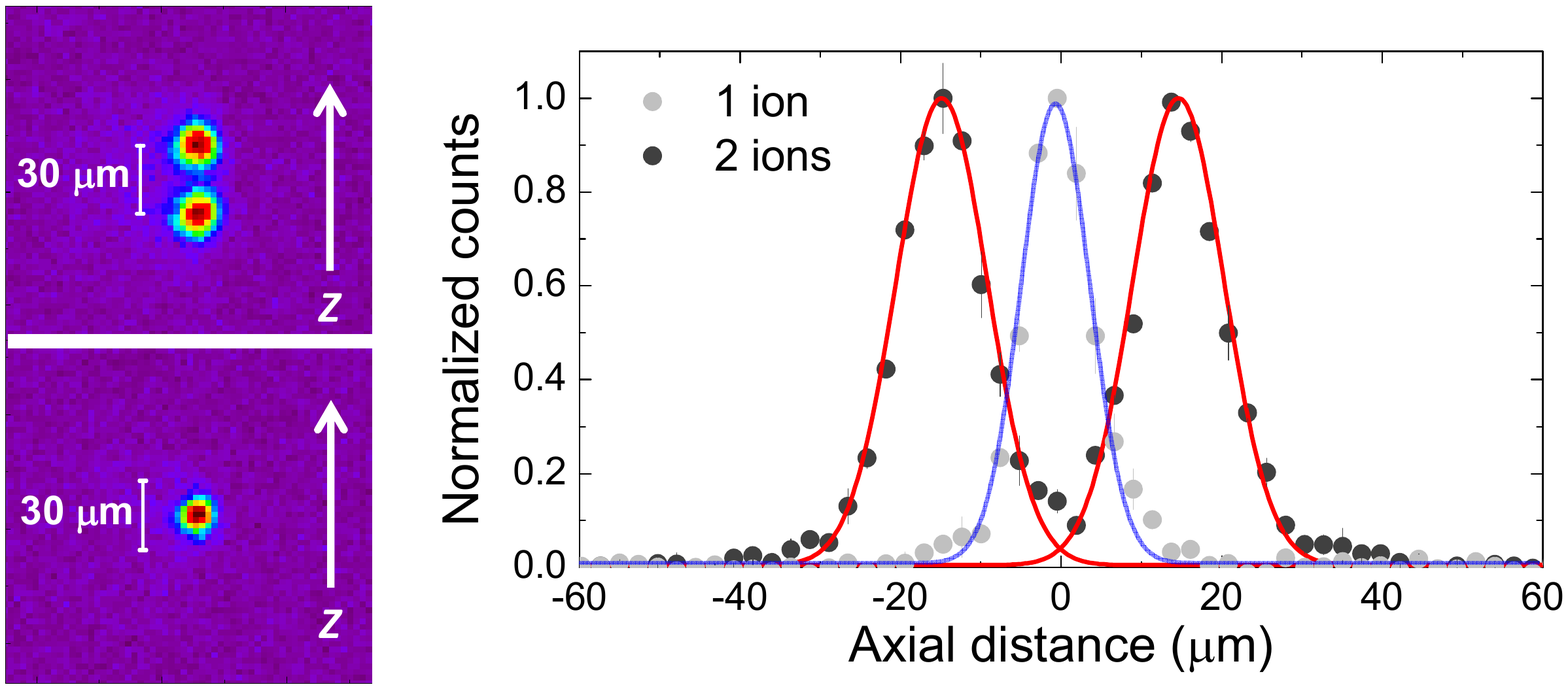}
\vspace{-5.5cm}
\caption{Fluorescence images (left) and projections in the axial direction (right) for one and two ions. The distance between the two-laser cooled ions is $30$~$\mu$m due to the balance between trapping potential and Coulomb repulsion.\label{fig_2}}
\end{figure}

The results presented here represent detailed investigations of the optical response of a two-ion crystal compared to the single ion case, for two types of external driving, namely harmonic dipolar and white noise fields (Fig.~\ref{fig_1}). The base line for these measurements is shown in Fig.~\ref{fig_2}, where one can observe the fluorescence image in the EMCCD sensor of one and two laser-cooled $^{40}$Ca$^+$ ions together with their axial distributions. A single ion is cooled down to a temperature below $10$~mK \cite{Domi2017} and performs a random walk around the equilibrium position in the axial potential well of the trap. When two ions are laser cooled below a certain temperature, they move around fixed average positions separated a distance \cite{Jame1998}

\begin{equation}
\Delta z=(e^2/2\pi \epsilon _0m\omega _{{\scriptsize{\hbox{z}}}}^2)^{1/3}\approx 30~\mu \hbox{m}, 
\end{equation} 
\noindent due to the balance between the Coulomb and trapping potentials. In this scenario, the centroids of the ion motions are $15$~$\mu$m apart from the minimum of the potential well. This originates some micromotion, so that the spread of the fluorescence distribution is not only caused by the finite resolution of the imaging system and the recoil heating. This can be seen from the comparison of the width of the individual distributions: $\sigma _{\scriptsize{\hbox{1-ion}}}^{\scriptsize{\hbox{single}}}=4.3$~$\mu $m versus $\sigma _{\scriptsize{\hbox{1-ion}}}^{\scriptsize{\hbox{crystal}}}=5.7$~$\mu $m.\\

\subsection{Harmonic dipolar field}

\begin{figure}[t]
\centering
\includegraphics[width=1.1\linewidth]{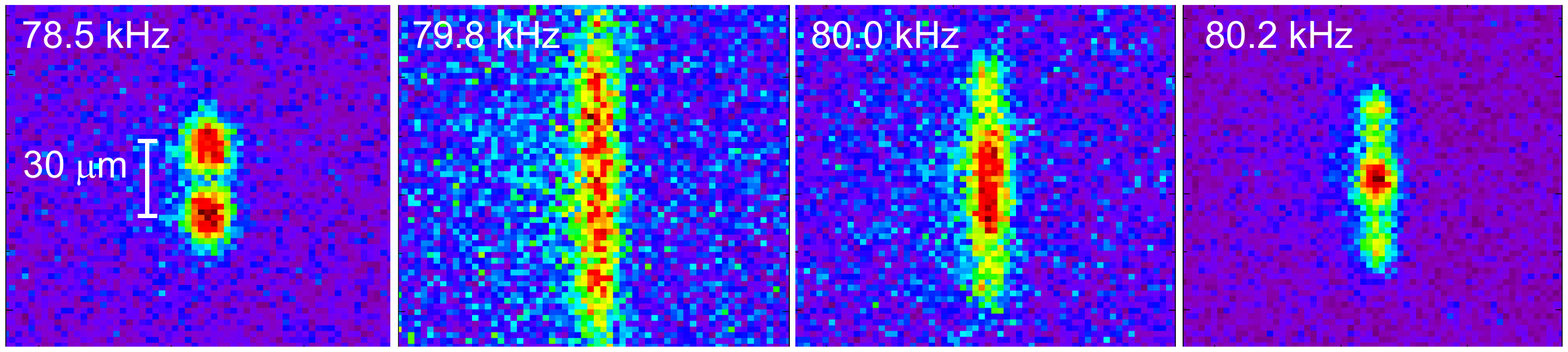}
\vspace{-8.1cm}
\caption{Fluorescence images of two laser-cooled ions when their COM mode is probed with an external dipolar field, driven at different frequencies close to resonance. The resonance frequency is $79.7$~kHz. \label{fig_3}}
\end{figure}

\begin{figure}[t]
\centering
\includegraphics[width=1.05\linewidth]{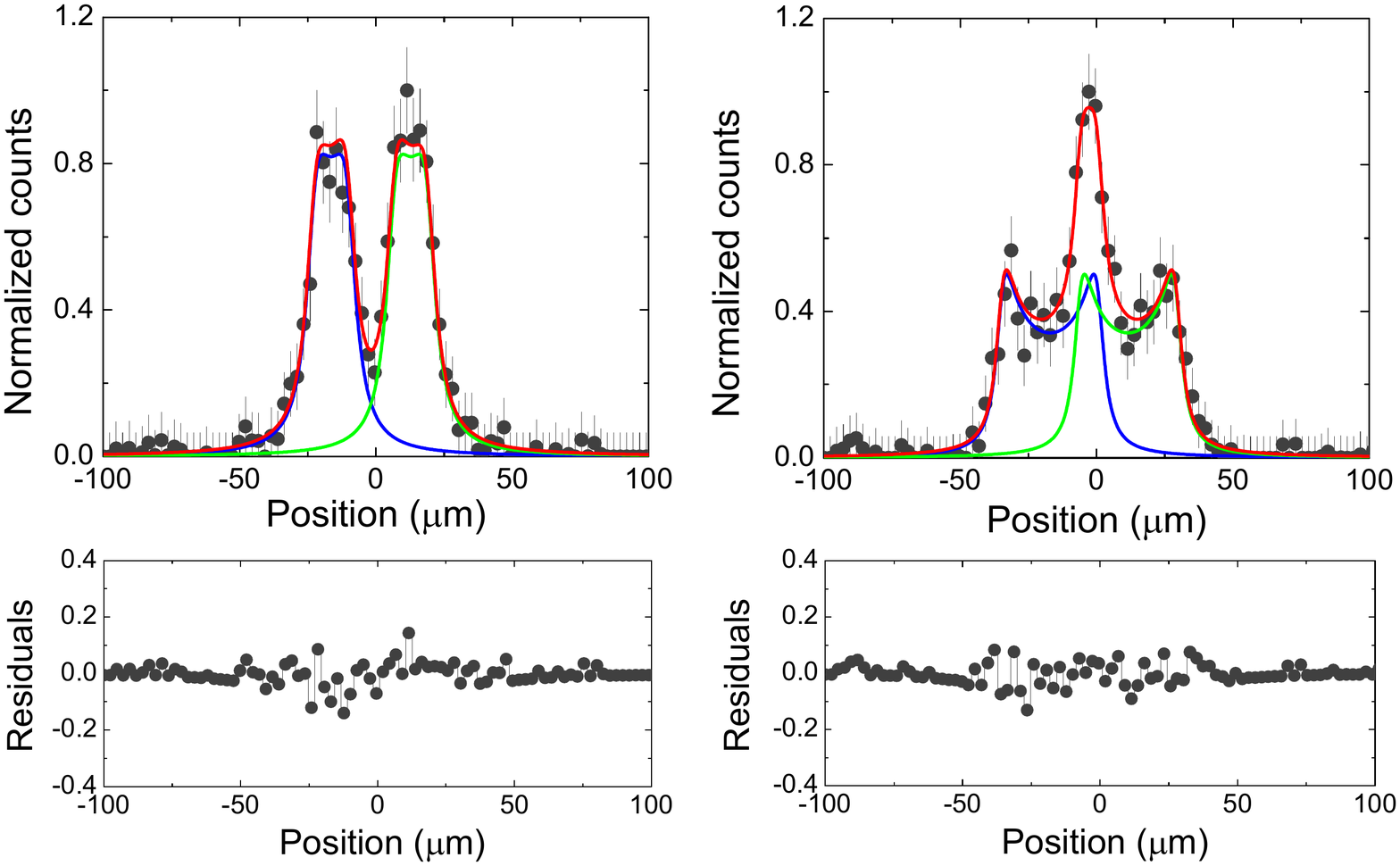}
\vspace{-2.4cm}
\caption{Axial photon distributions for two of the fluorescence images shown in Fig.~\ref{fig_3}. The frequencies of the dipolar field are 78.5~kHz (left) and 80.2~kHz (right). The uncertainty for each data point corresponds to 3\,$\sigma$. The solid lines represent the fits following Eq.~(\ref{fitting}). The lower panels show the residuals from each fit. \label{fig_7}}
\end{figure}

The response of a single ion to a harmonic dipolar force in resonance with the ion's motion has been recently studied \cite{Domi2017}. Here, we compare this response with that of a two-ion crystal. Two ions confined in the same potential well and driven by a harmonic field of intensity above the noise level oscillate in phase following the driving force. Figure~\ref{fig_3} shows the fluorescence image of two laser-cooled ions probed at different frequencies $\omega _{\scriptsize{\hbox{dip}}}$ of the dipolar field. At $\omega _{\scriptsize{\hbox{dip}}}=2\pi \times 78.5$~kHz, the oscillation amplitude of the ions in the steady state is below half the distance between them, and therefore they can be individually observed. Increasing $\omega _{\scriptsize{\hbox{dip}}}$ towards $\omega _{\scriptsize{\hbox{z}}}$, the oscillation amplitude increases until the fluorescence distributions of the two ions start to overlap, leading to a characteristic image. The axial distribution spreads further approaching the resonance and decreases again thereafter (when $\omega _{\scriptsize{\hbox{dip}}}>\omega _{\scriptsize{\hbox{z}}}$), as observed for $\omega _{\scriptsize{\hbox{dip}}}=2\pi \times 79.8$~kHz, $2\pi \times 80.0$~kHz and $2\pi \times 80.2$~kHz. In the latest case, the oscillation amplitude is approximately half of the separation distance between the ions. Projections into the axial plane for two of those images are shown in Fig.~\ref{fig_7}, along with the fits to the model discussed in the following.\\ 

\noindent The image obtained on the EMCCD is a convolution of the time-averaged motion of the ion with the resolution of the optical system. Therefore, the fits to the experimental data in Fig.~\ref{fig_7} are performed using a function resulting from the convolution of a Lorentzian with the probability density function of the harmonic oscillator as presented in Ref.~\cite{Domi2017}, but now using the analytical solution of the following integral \cite{wesenberg2007}:
\begin{equation}
	F(z)=\int \frac{\Gamma}{2\pi^2}\frac{1}{\left(\frac{\Gamma}{2}\right)^2+(f-(x-x_0))^2}\sqrt{\frac{1}{\rho _{\scriptsize{\hbox{z,max}}}^2-f^2}}df,
	\label{eq:int}
\end{equation}
\noindent where $f$ is the integration variable, $\Gamma$ is the Lorentzian width accounting for the resolution of the optical system taking into account the motion of the ion, $z$ is the position of observation, and $z_0$ is the equilibrium position of the ion. The analytical solution is given by \cite{wesenberg2007}:
\begin{equation}
	F(z)=\frac{2}{\pi\Gamma^2}\frac{A_0}{2\rho _{\scriptsize{\hbox{z,max}}}} \text{Im}\left[\frac{i}{\sqrt{1-\Gamma^2\frac{\left(\frac{2(z-z_0)}{\Gamma}+i\right)^2}{4\rho _{\scriptsize{\hbox{z,max}}}^2}}}\right], \label{fitting}
\end{equation}
\noindent where $\rm Im$ stands for the imaginary part. The results from the fits to each of the ion's distribution following Eq.~(\ref{fitting}), $F(z_1)$, $F(z_2)$ and the sum $F(z_1)+F(z_2)$, are also shown in Fig.~\ref{fig_7}. There are five free parameters: the width of the Lorentzian distribution, the central position for each ion $z_1$ and $z_2$, with $z_0=(z_2-z_1)/2$, $\rho _{\scriptsize{\hbox{z,max}}}$($=\rho _{\scriptsize{\hbox{z,max}}}^{\scriptsize{\hbox{ion1}}}=\rho _{\scriptsize{\hbox{z,max}}}^{\scriptsize{\hbox{ion2}}}$), and a scaling parameter $A_0$. We have observed that the Lorentizian width might vary for large amplitudes of $\rho _{\scriptsize{\hbox{z,max}}}$ and we assign this to the effect of the micromotion, which has an amplitude of about 0.10-0.15$\rho _{\scriptsize{\hbox{z,max}}}$. 

\subsection{Sensitivity: one and two ions}

\begin{figure}[ht]
\centering
\includegraphics[width=0.8\linewidth]{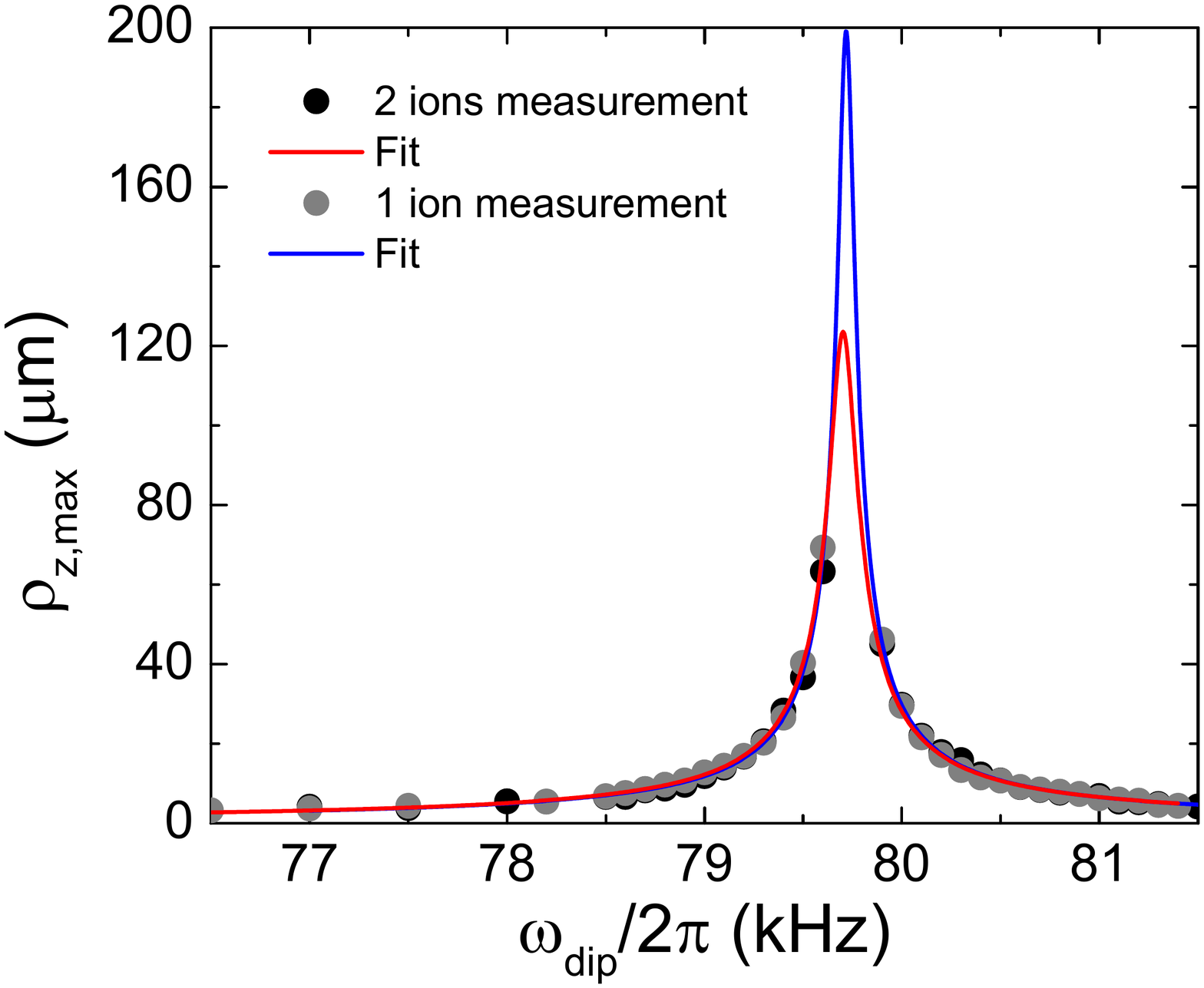}
\vspace{0cm}
\caption{Oscillation amplitude of one ion as a function of the frequency of the dipolar field. The solid-lines represent the fits following Eq.~(\ref{fit_osc}). Two situations have been considered: amplitude of a single ion, and the amplitude of one ion, which is part of a two-ion crystal. \label{fig_8}}
\end{figure}

The oscillation amplitude of the driven ions $\rho _{\scriptsize{\hbox{z,max}}}$  is a measure of the sensitivity of the system to an external force. This is obtained using the function given by Eq.~(\ref{fitting}). Figure~\ref{fig_8} shows $\rho _{\scriptsize{\hbox{z,max}}}$ as a function of the frequency of the external dipolar field ($\omega _{\scriptsize{\hbox{dip}}}/2\pi$) for one and two ions. The resulting data are fitted using the steady state solution of a driven-damped harmonic oscillator \cite{Domi2017}

\begin{equation}
\rho _{z,\scriptsize{\hbox{max}}}=\frac{F_e}{m} \{ (2 \gamma_z \omega_{\rm dip})^2 + (\omega_z^2 - \omega_{\rm dip}^2)^2 \}^{-1/2}, \label{fit_osc}
\end{equation}

\noindent where $F_e$ and $\omega_{\rm dip}$ are, respectively, the amplitude and the frequency of the harmonic driving force, and $\gamma_z$ the damping coefficient. In the single ion case $m$ corresponds to the ion’s mass, whereas in the two-ion case $m$ represents the mass of the centre-of-mass motion. $F_e$ is the same for both, single ion and two-ion crystal and the result is in agreement with previous results quoted in \cite{Domi2017}. An outcome from the measurements is that the damping coefficient ($\gamma _z$) differs: $\gamma _z=309(21)$~s$^{-1}$ for a single ion, against $\gamma _z=354(20)$~s$^{-1}$ for two ions. This can be observed in Fig.~\ref{fig_8}, where the solid lines are the fitting curves. Assuming the same laser conditions, this difference can be assigned to the Coulomb interaction between the two ions, which causes stronger damping. $F_e$ is the same for both, single ion and two-ion crystal, and the result is in agreement with previous results quoted in \cite{Domi2017}. 

\subsection{White noise}

\begin{figure}[b]
\centering
\includegraphics[width=1.15\linewidth]{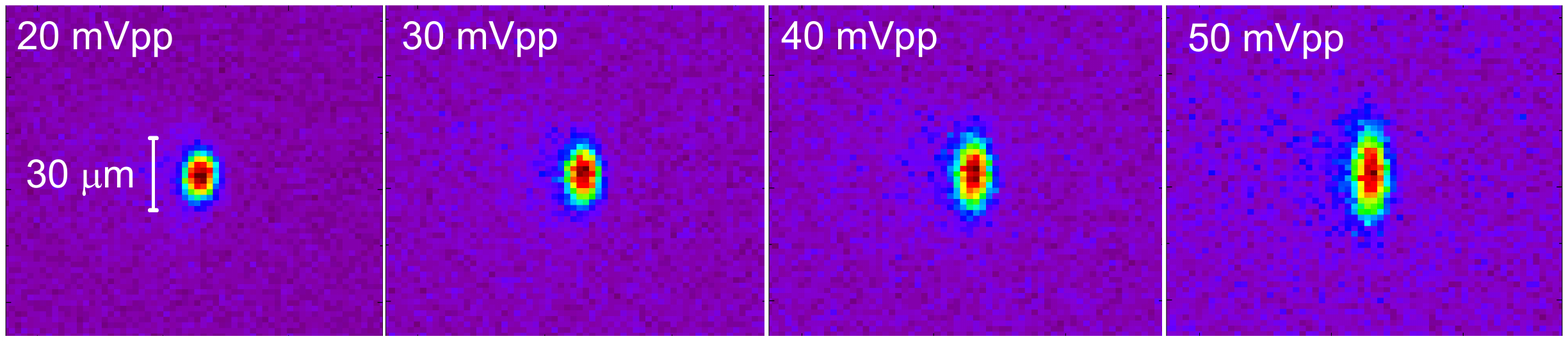}
\vspace{-8.5cm}
\caption{Evolution of the fluorescence image of a single laser-cooled ion for different amplitudes of an applied noise signal.\label{fig_4}}
\end{figure}

\begin{figure}[ht]
\centering
\includegraphics[width=1.1\linewidth]{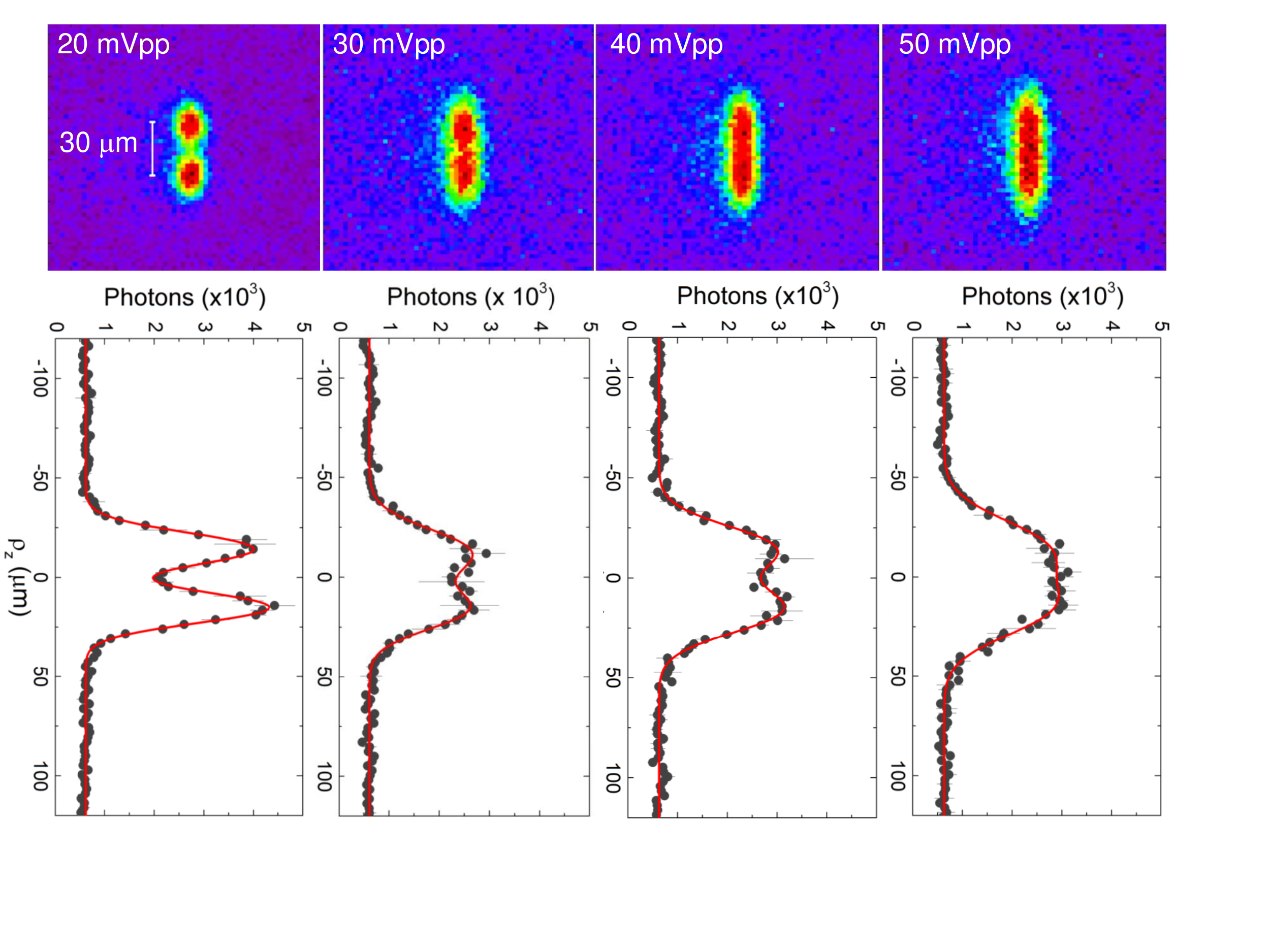}
\vspace{-1.8cm}
\caption{Top: Evolution of the fluorescence image of two laser-cooled ions with increasing amplitudes of an applied noise signal. Bottom: Axial photons distributions.\label{fig_5}}
\end{figure}

Electrical noise at the surfaces of trap electrodes leads to heating of the trapped ions~\cite{noise2015RMP}. Although this is an unwanted effect, it has been shown that a well-controlled external noisy force acting on the ions can provide valuable information about their dynamics, in particular on their statistical and thermodynamic descriptions~\cite{Mart2016,rossnagel2016single,ricci2017optically}. We have studied the dynamic response of one and two laser-cooled ions to white noise applied to the electrode, as shown in Fig.~\ref{fig_1}. Figures~\ref{fig_4} and \ref{fig_5} show the fluorescence images for different amplitudes of the white noise for one and two ions, respectively. We see that increasing levels of noise lead to wider distributions of the position, due to enhanced Brownian motion. Note that this enhancement only takes place in the axial direction, parallel to the applied field. The width, and hence the temperature of the radial modes remain undisturbed, since there is no coupling between modes. The noise leads to an additional heating rate given by \cite{noise2015RMP}
 
\begin{equation}
	\Gamma_h\simeq\frac{e^2}{4m\hbar \omega _{\scriptsize{\hbox{z}}}}S_E(\omega_{\scriptsize{\hbox{z}}}),
	\label{eq:heatingrate}
\end{equation}

\noindent where $S_E(\omega_{\scriptsize{\hbox{z}}})$ is the single-sided spectral density of the noise at frequency $\omega_{\scriptsize{\hbox{z}}}$. Considering a random signal characterized by a Gaussian white noise process, $S_E(\omega_{\scriptsize{\hbox{z}}})$ is proportional to the intensity of the electric--field fluctuations and thus $S_E(\omega_{\scriptsize{\hbox{z}}})\propto V_{\rm noise}^2$. This additional heating source adds up to the natural heating rate in the laser-cooling process due to photon recoil, defining a new Doppler limit, which in the simplest case can be written as \cite{Domi2017}:
\begin{equation}
	T_{\scriptsize{\hbox{z,Doppler}}}=\frac{1}{\gamma _z k_B}(K+\zeta \cdot V_{\scriptsize{\hbox{noise}}}^2),
	\label{eq:TDoppler}
\end{equation}
where $\gamma_z$ is the decay time constant (damping coefficient) in the cooling process, $K$ is a constant, and $\zeta \cdot V_{\scriptsize{\hbox{noise}}}^2$ is the heating rate in units of J$\cdot$ s$^{-1}$. Of course, we recover the standard Doppler limit when $V_{\scriptsize{\hbox{noise}}} =0$.  Therefore, we expect a linear relation between the temperature of the ion and the intensity of the white noise applied to the electrodes. \\

\begin{figure}[t]
\begin{center}
\includegraphics[width=0.8\textwidth]{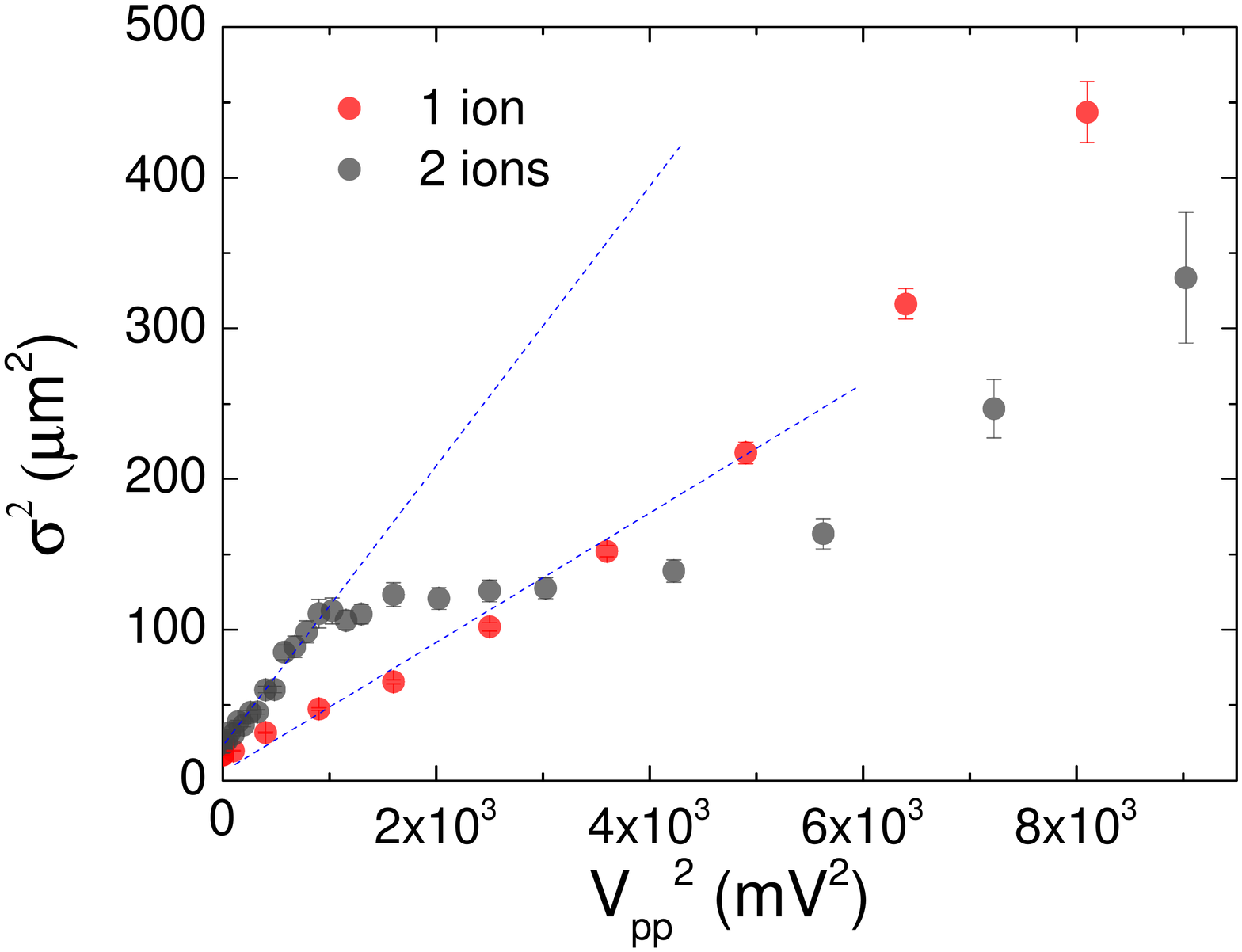}
\caption{Standard deviation ($\sigma^2$) obtained from the axial projection of ion-fluorescence image versus noise intensity ($\hbox{V}_{\scriptsize{\hbox{pp}}}^2$). The data points correspond to one ion, when it is isolated (red circles) or it is part of a two-ion crystal (grey circles). The blue dashed lines are used to guide the eyes.}
\label{fig_6}
\end{center}
\end{figure}

\noindent Figure~\ref{fig_6} shows  the effects of white noise on the width of the fluorescence distributions. This is the outcome from the analysis of the measurements partially shown in Figs.~\ref{fig_4} and \ref{fig_5}. For a single ion, the linear relation given by Eq.~(\ref{eq:TDoppler}) holds true. Note that Equipartition theorem sets $T_z \propto \sigma ^2$. This linear dependence breaks down for high intensities of the field and high motional amplitudes, as expected due to the simplifications made in the derivation of Eq.~(\ref{eq:TDoppler})~\cite{Blat2003}. At high intensities, and therefore high amplitudes of motion, micromotion is increasingly more relevant, leading to a higher heating rate. The linear dependence is also observed for two ions at low intensities of the noise signal. Interestingly, this linear region is followed by a plateau in the value of $\sigma^2$, in the interval of $V_{\rm noise}^2\in[1000-4000]$~$\rm mV^2$. After this plateau, $\sigma^2$ increases again with $V_{\rm noise}$, but with a smaller slope, and thus with smaller heating rate, reaching a value similar to that of the single ion at those amplitudes. \\

\noindent In our experimental situation, the external noise is perfectly spatially correlated. In such a case, the heating rate for the COM mode of the crystal is proportional to the number of ions $N$ in the string, i.e. $\Gamma_h{(\hbox{COM})}=N\Gamma_h$ \cite{noise2015RMP}, so that

\begin{equation}
	\Gamma _{h\rm 2 ions}{(\hbox{COM})}\simeq\frac{e^2}{2m\hbar\omega_{\rm COM}}S_E(\omega_{\rm COM}),
	\label{eq:heatingratetot}
\end{equation}
\noindent and therefore the COM mode of the two-ion crystal absorbs more energy from the field than an isolated ion, resulting in a higher heating rate for each individual ion. In fact, the initial slope for the ion string is almost twice of that for the single ion. The beginning of the plateau coincides with the value of the noise at which the two lobes in the fluorescence distribution (Fig.~\ref{fig_5}) are not distinguishable anymore and they merge into a single one. This can be interpreted as the melting of the Coulomb crystal, or an order-to-chaos transition, expected to occur when the kinetic energy of the ions surpasses the Coulomb potential~\cite{Thom2015}. The presence of the plateau suggests there is a region of coexistence of two phases, the ion crystal and the melted cloud. Since the heating rate for the melted cloud is lower than that for the crystal, it might happen that the two ions melt and recrystallize in the region where the kinetic and repulsive energies are similar. In this situation, fluctuations are responsible for the jumps between these two states. In the images taken with the EMCCD camera (see Fig.~\ref{fig_5}), one can observe an average over crystal and melted cloud, blurring the visualization of the two Gaussian distributions. Increasing the intensity of the noise field, the kinetic energy eventually overcomes the Coulomb repulsion and the ions heat with the heating rate of the cloud phase. In this phase, the correlation between the two ions is low, they behave like two isolated particles, thus experiencing a similar heating rate to the single-ion case. 

\section{Outlook and perspectives: projection to Penning traps}

In this publication, we have described a fully-developed optical detection method that will allow determining with precision the oscillation amplitude of the motion of one or two ions confined in a trap, when they are laser-cooled to temperatures in the order of several mK. We have also studied the response of trapped ions to an external and controllable source of noise, which is important for experiments where one ion is studied through the fluorescence emitted by the neighbour one, both confined in the same potential well or in different traps.\\

\noindent In the next stage of the experiment, which is headed towards the realization of the so-called \textit{Quantum-Sensor} \cite{Rodr2012, Corn2012} technique by means of a double micro Penning-trap system \cite{Corn2016}, the method of optical ion sensing will be applied to a two-ion crystal  in a Penning trap to carry out Sympathetically-Cooled Single Ion Mass Spectrometry (SCSI-MS)\cite{Staa2010}. The crystal will consist of a single $^{40}$Ca${^+}$ ion of mass $m$, and a singly-charged heavy or super-heavy ion (e.\,g. $^{163}$Ho, $^{187}$Re, $^{187}$Os) of mass $M$, stored along the magnetic field. The SCSI-MS method relies on the determination of the mass ratio $\mu=M/m$, by measuring the eigenfrequencies of the crystal in the harmonic potential well of the trap. According to \cite{Mori2001}, the modified motional frequencies $\Omega_{\pm}$ for a two-ion Coulomb crystal reads
\begin{align}
\Omega_{\pm}^2=\omega_{z,\,\mu=1}^2\left(1+\frac{1}{\mu}\pm\sqrt{1+\frac{1}{\mu^2}-\frac{1}{\mu}}\right),
\label{Eq.1}
\end{align}
where $\omega_{z,\,\mu=1}=\omega_{\text{ref}}$ represents the COM eigenfrequency of two ions of equal mass. 
It has been shown in a linear Paul trap \cite{Drew2004}, that this method provides only a mass resolution $M/\Delta M$ of $\sim 10^{2}$- $10^{3}$, due to systematic effects, like shifts in the COM eigenfrequency originated by damping effects introduced by the laser-ion interaction, or instabilities in the trapping potential. However, using this scheme in a Penning trap might allow improving $M/\Delta M$, and also extending the applicability to heavy and superheavy elements, opposite to Paul traps, where there is a limitation arising from $0.5\lesssim  \mu \lesssim 2$. Furthermore, the harmonicity of the trapping potential in a Penning trap is often orders of magnitude better compared to a Paul trap, since there are ultra-stable DC voltage sources \cite{Boeh2016}, which provide the potentials with an accuracy of 1~ppm.\\

\begin{figure}[t]
\centering
\includegraphics[width=0.8\linewidth]{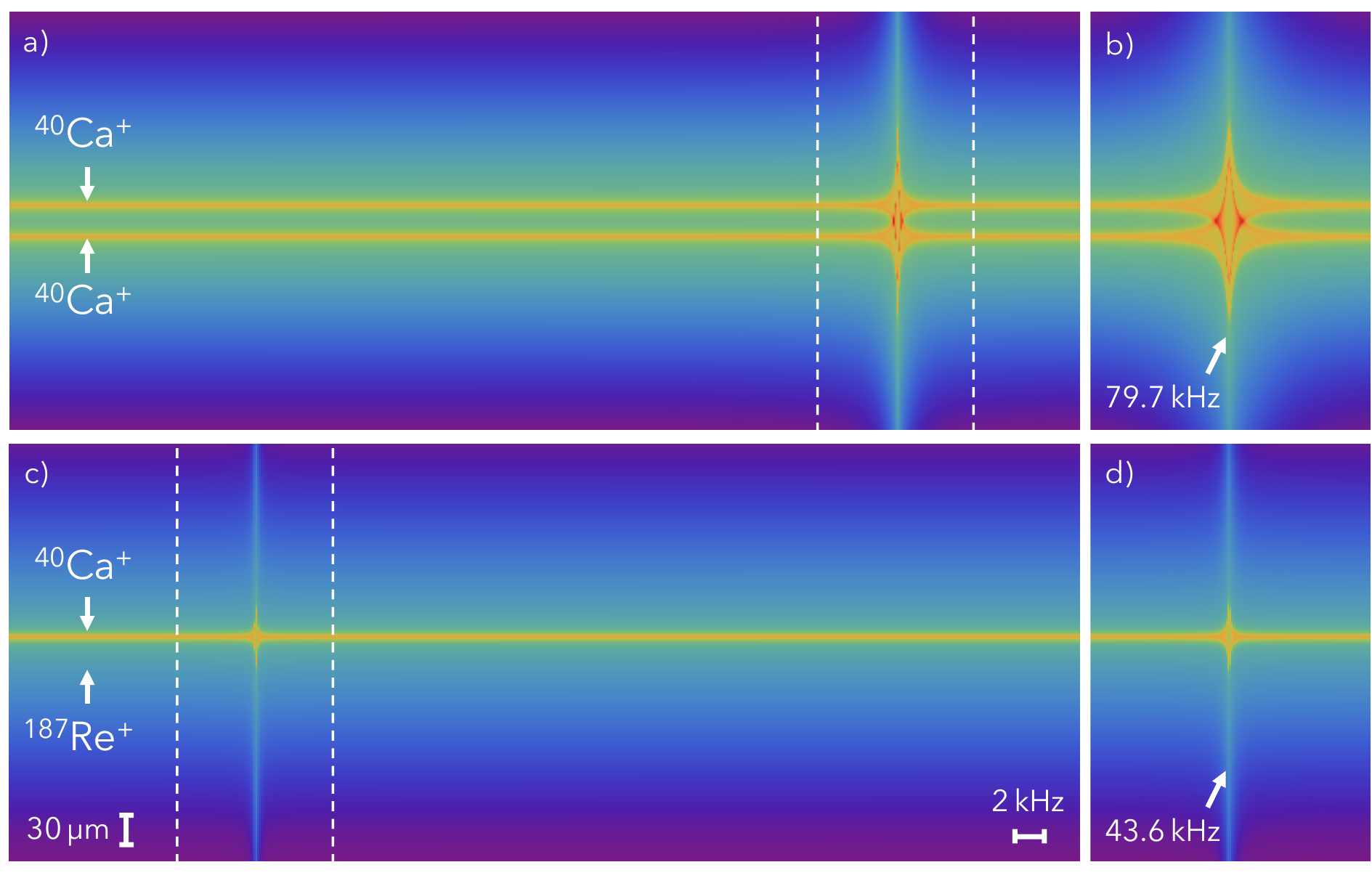}
\vspace{0cm}
\caption{Motional spectra of a two-ion crystal as a function of the dipolar frequency. The spectra are modelled based on the results obtained in this work. According to Eq.~(\ref{Eq.1}), the predicted eigenfrequency $\Omega_{-}$ of the two-ion crystal $^{40}$Ca$^{+}$\,--\,$^{187}$Re$^{+}$ (c) is shifted to a lower frequency with respect to the COM frequency of $^{40}$Ca$^{+}$\,--\,$^{40}$Ca$^{+}$ (a). (b) and (d) are enlarged representations of the areas within the vertical dashed lines in (a) and (c), respectively. \label{fig_9}}
\end{figure}

\noindent Based on the results shown in Figs.~\ref{fig_7} and \ref{fig_8},  Fig.\,\ref{fig_9} illustrates the expected axial fluorescence distribution of a two-ion crystal $^{40}$Ca$^{+}$\,--\,$^{187}$Re$^{+}$ in a Penning-trap, with $\omega_{\text{ref}}$ set to 79.7\,kHz, a resolution of the imaging system of 6\,$\mu$m (Lorentzian width) and an amplitude for the dipolar field of 1\,mV$_{\scriptsize{\hbox{pp}}}$.  The mass of the dark $^{187}$Re$^{+}$ ion can be obtained from a precise determination of one of the eigenfrequencies $\Omega_{\pm}$ of the two-ion crystal. \\

\noindent In the Penning trap, further improvement in the mass resolving power, up to a level of $\sim 10^{-6}$-$ 10^{-8}$ can be obtained from a direct measurement of the cyclotron frequency $\omega_{c}=qB/m$ of the dark ion, combining the high-sensitivity of the optical method with the well-established Time-Of-Flight  Ion-Cyclotron-Resonance (TOF-ICR) technique \cite{Koen1995}, and later by removing the Coulomb interaction if the reference ion and the ion of interest are confined in different traps \cite{Rodr2012}. Both methods rely on controlled energy transfer between the radial and the axial motional degrees of freedom of the two ions, since the detection is done through the axial motion of the $^{40}$Ca$^+$ ion.\\

\noindent It has been shown recently, that the highest sensitivity in a Penning trap reaches a level of up to 50\,pm \cite{Gilm2017}. Although this is about 5 to 6 orders of magnitude better than the results presented here, the scheme proposed in this work, i.e., optical monitoring of the ion motion, is highly valuable when short measurement times, as well as non-destructive measurements with single-ion sensitivity, are mandatory. Improving the resolution of the optical system and increasing the magnification will improve sensitivity within the same measurement times. This might be enhanced if a cooling mechanism allowing reaching sub-mK temperatures is implemented.

\section*{Acknowledgement(s)}



We acknowledge support from the European Research Council (ERC StG contract 278648-TRAPSENSOR); Spanish MINECO/FEDER FPA2012-32076, FPA2015-67694-P, UNGR10-1E-501,  UNGR13-1E-1830,  FIS2015-69983-P; Junta de Andaluc\'ia/FEDER IE-57131; Basque Government PhD grant PRE-2015-1-0394 and project IT986-16. F.D. acknowledges support from Spanish MINECO ``Programa de Garant\'ia Juvenil'' cofunded by the University of Granada. M.J.G.'s  work  is  supported by the Spanish MECD through the grant FPU15-04679. R.A.R. acknowledges support from the "Juan de la Cierva-Incoporaci\'on" program from MINECO.  S.S and J.J.D.P. acknowledges support from the University of Granada thorugh the ``Plan propio (Programa de Intensificaci\'on de la Investigaci\'on)".

\end{document}